\documentclass[runningheads]{llncs}
\usepackage{graphicx}
\usepackage{xcolor}
\usepackage{multirow}

\begin{document}
\title{Interoperability and Integration Testing Methods for IoT Systems: a Systematic Mapping Study}

\titlerunning{Interoperability and Integration Testing Methods for IoT Systems}
%
\author{Miroslav Bures\inst{1}\orcidID{0000-0002-2994-7826} \and
Matej Klima\inst{1}\orcidID{0000-0002-9601-8787} \and
Vaclav Rechtberger\inst{1}\orcidID{0000-0002-4127-5372} \and
Xavier Bellekens\inst{2}\orcidID{0000-0003-1849-5788} \and
Christos Tachtatzis\inst{2}\orcidID{0000-0001-9150-6805} \and
Robert Atkinson\inst{2}\orcidID{0000-0002-6206-2229} \and
Bestoun S. Ahmed\inst{1,3}\orcidID{0000-0001-9051-7609}}
\authorrunning{M. Bures et al.}
%
\institute{Dept. of Computer Science, FEE, Czech Technical University in Prague, Karlovo namesti 13, , Czech Republic \\
\email{miroslav.bures@fel.cvut.cz}\\
\url{http://still.felk.cvut.cz}
\and
Dept. of Electronic and Electrical Engineering, University of Strathclyde, 99 George Street, Glasgow, Scotland, UK
\and
Dept. of Mathematics and Computer Science, Karlstad University, Universitetsgatan 2, 65188, Karlstad, Sweden}
\maketitle              
\begin{abstract}
The recent active development of Internet of Things (IoT) solutions in various domains has led to an increased demand for security, safety, and reliability of these systems. Security and data privacy are currently the most frequently discussed topics; however, other reliability aspects also need to be focused on to maintain smooth and safe operation of IoT systems. Until now, there has been no systematic mapping study dedicated to the topic of interoperability and integration testing of IoT systems specifically; therefore, we present such an overview in this study. We analyze 803 papers from four major primary databases and perform detailed assessment and quality check to find 115 relevant papers. In addition, recently published testing techniques and approaches are analyzed and classified; the challenges and limitations in the field are also identified and discussed. Research trends related to publication time, active researchers, and publication media are presented in this study. The results suggest that studies mainly focus only on general testing methods, which can be applied to integration and interoperability testing of IoT systems; thus, there are research opportunities to develop additional testing methods focused specifically on IoT systems, so that they are more effective in the IoT context.

\end{abstract}

\section{Introduction}
\label{sec:introduction}

\color{blue}\textbf{Accepted at the Software Engineering and Formal Methods (SEFM) 2020, Amsterdam, September 14--18th, https://event.cwi.nl/sefm2020/}\color{black}

The Internet of Things (IoT) provides numerous advantages to its users in various application domains. However, extensive development of IoT systems in the last decade has led to a number of reliability and security challenges \cite{bures2018internet,kiruthika2015software,khan2018iot}. One of the challenges frequently reported by researchers as well as industry practitioners is the integration testing of IoT systems. In contemporary IoT projects, software developers work with network specialists and electronic experts for testing; however, these parties have different backgrounds and may be accustomed to using different methods of system testing (e.g., low-level testing vs. high-level functional testing). Moreover, different expectations may also play a role; for example, in a standard software system, lower layers (e.g., network or operating systems) are usually considered to be already tested and reliable; therefore, quality engineers focus on the application itself. In the case of an IoT system, the situation might differ and lower levels might also need to be tested properly. 
In addition, interoperability challenges are closely associated with integration testing; different devices using a variety of protocols need to cooperate in an appropriate manner, and this reliable cooperation has to be verified. Individual devices can have numerous versions and variants, which increases the difficulty of correct and seamless integration.

Integration testing and interoperability testing of IoT systems are considered to overlap for several cases even though semantic differences and different definitions can be pointed out. However, because these terms overlap in their common usage, we decided to cover both \textbf{interoperability} and \textbf{integration} testing in the scope of this study.

As mentioned earlier, there is an increased demand for more efficient interoperability and integration testing methods. Currently, the model-based testing (MBT) discipline naturally covers the area of integration testing through methods such as path-based testing \cite{bures2019employment,anand2013orchestrated,bures2015pctgen}, which is typical for E2E integration tests, or combinatorial \cite{nie2011survey} and constrained interaction testing \cite{ahmed2017constrained}, which is useful in unit integration testing and optimization of system configurations. Logically, in the recent period, researchers have attempted to tailor or apply formal verification and MBT techniques for IoT systems to increase system efficiency \cite{ahmed2019aspects}.
Interoperability and integration testing have significant importance in the IoT context and mapping current methods for IoT integration testing would provide valuable information to researchers for IoT and industrial quality assurance. Unfortunately, no systematic mapping study has been conducted yet in the field of integration testing for IoT systems. Hence, we attempt to bridge this gap through this study.

The contributions of this study are as follows:
\begin{enumerate}
\item It gives an overview of research and development activity in this field, identifying the active parties and individuals;
\item It also provides an overview of methods and approaches that are available for IoT integration testing;
\item It identifies research opportunities and discusses possible research directions.
\end{enumerate}

This paper is organized as follows. Section \ref{sec:motivation} analyzes existing mapping studies and literature surveys in the fields of integration testing, IoT testing, quality assurance and IoT integration, which justifies the motivation of this study. Section \ref{sec:methodology} explains the methodology used in this study, defines the research questions (RQs) to be answered, and the stages through which relevant studies are identified and analyzed. Section \ref{sec:results} presents the answers to individual RQs and related discussions. The last section presents the analysis of the possible threats to the validity of this study and concludes the paper.

\section{Motivation and Related Work}
\label{sec:motivation}

The motivation of this study is twofold. The first is the importance of integration testing in the quality assurance process of IoT solutions \cite{bures2018internet,kiruthika2015software} and the second is the fact that no previous systematic mapping study has addressed integration testing methods for IoT systems specifically.

In the field of general integration testing, there are several systematic literature surveys and mapping studies.

In 2007, Rehman et al. published a survey of issues and available techniques for general software integration testing \cite{jaffar2007testing}. Their study summarizes and classifies a variety of integration testing approaches, covering the fields of MBT, test automation frameworks, and methodological aspects; it also provides a good overview of available approaches and concepts that can be used in the definition of a test strategy. However, the study focuses on general software integration testing and is not IoT-specific. Moreover, the study was published more than a decade ago; new techniques and approaches might be available now. Moreover, modern integrated software applications may change as the systems are becoming more complex and demands for their real-time or almost real-time operation have increased. This will also be reflected in integration testing methods; therefore, a state-of-the-art survey is required. 

A more recent study by Shashank et al. from 2010 also focuses on the field of integration testing of component-based software systems. However, the study, published as a conference paper, is limited in terms of its sample size; rather than an extensive classification, it provides an overview of available approaches and selected examples of approaches \cite{shashank2010systematic}. Despite the limited extent of the study, the brief classification of the state-of-the-art methods into established MBT and software verification categories provided in this study is valid.

Another recent survey and analysis on model-based integration testing was conducted by Haser et al. in 2014 \cite{haser2014software}. Essentially, this study is not limited to software systems; the authors discuss integration testing methods that can be applied to a broader scope of cyber-physical systems, which also covers the IoT domain. In the study, an extensive sample of 718 papers is analyzed, and conclusions are obtained for the defined research questions on software paradigms, system assessment types, and usage of non-functional requirements. However, the study is limited to model-based integration testing with limited scope of defined research questions. For the field of IoT-specific integration testing methods, a broader study is required.

In the field of testing techniques that specifically focus on IoT systems and their specifics, a recent systematic mapping study by Ahmed et al. \cite{ahmed2019aspects} focuses on general aspects of quality and quality assurance techniques designed for IoT systems. The scope of this study is broader than the field of integration testing and covers topics such as security, privacy, construction of testbeds, general MBT, and formal verification techniques. Integration testing is not discussed in depth in this study due to its general scope, and from this viewpoint, overlap with the scope of this study is minimal.

Another recent conference paper by Dias et al. briefly summarizes current testing tools for IoT systems; integration testing is included in the examined aspects of the problem \cite{dias2018brief}. However, the discussion is brief, and regarding the selected method in the study, all state-of-the-art methods in this field are not covered.

In 2019, Cortes et al. conducted a mapping study on software testing methods used for IoT systems \cite{cortes2019adoption}. The study categorizes and analyses publications discussing general testing approaches used in IoT systems. Unfortunately, the discussion of integration testing is very brief in this paper.

Another study by Garousi et al. focuses on the testing methods for embedded systems \cite{Garousi_2018} (which may, to a certain extent, overlap with IoT systems discussed in this study). However, besides the fact that the field of embedded systems is not the same as the IoT field, the study focuses on general testing methods and approaches and does not concentrate on interoperability and integration testing specifically.

The most frequently addressed quality aspects of IoT systems in the last five years are security and privacy \cite{ahmed2019aspects}. This is also clear from the availability of published literature surveys and systematic mapping studies. A meta-survey summarizing and analyzing 32 available surveys on security, privacy, and defensive mechanisms of cyber-physical systems (including IoT) was recently published by Giraldo et al. \cite{giraldo2017security}. The study provides a good overview of previous works and motivates the reader to find relevant literature sources related to security and privacy problems.

Regarding the integration of IoT systems, a mapping study focusing on integration techniques and styles as well as related architectural aspects of integration was published by Cavalcante et al. \cite{cavalcante2016interplay}. However, this study does not discuss testing or quality assurance aspects of system integration.

To summarize, no current systematic mapping study is dedicated to integration testing techniques for IoT systems, discussing these techniques in the context of IoT domain and from the viewpoint of IoT quality challenges, which are frequent subjects of various reports\cite{bures2018internet,kiruthika2015software,khan2018iot}. This study aims to provide the missing information in this specific field.






\section{Methodology}
\label{sec:methodology}

This systematic mapping study follows the methodology recommendations provided by Kitchenham and Charters \cite{kitchenham2007guidelines}. The process of collection and analysis of relevant studies is divided into the following six stages:

\begin{enumerate}
\item Research scope determination and definition of RQs to be answered in the study.
\item Search for potentially relevant papers, which includes establishment of a search strategy and acquisition of the identified papers.
\item Identification of truly relevant papers from the initial selection based on the title, abstract, full-text, and quality assessments, which includes performing snowball sampling of other relevant studies.
\item Data extraction from the remaining papers to allow further detailed analyses.
\item Classification of papers and analyses of the extracted data to answer defined RQs.
\item Validity evaluation and discussion of the possible limitations of the study.
\end{enumerate}

\begin{figure*}[htbp]
\centering
  \includegraphics[width=12cm] {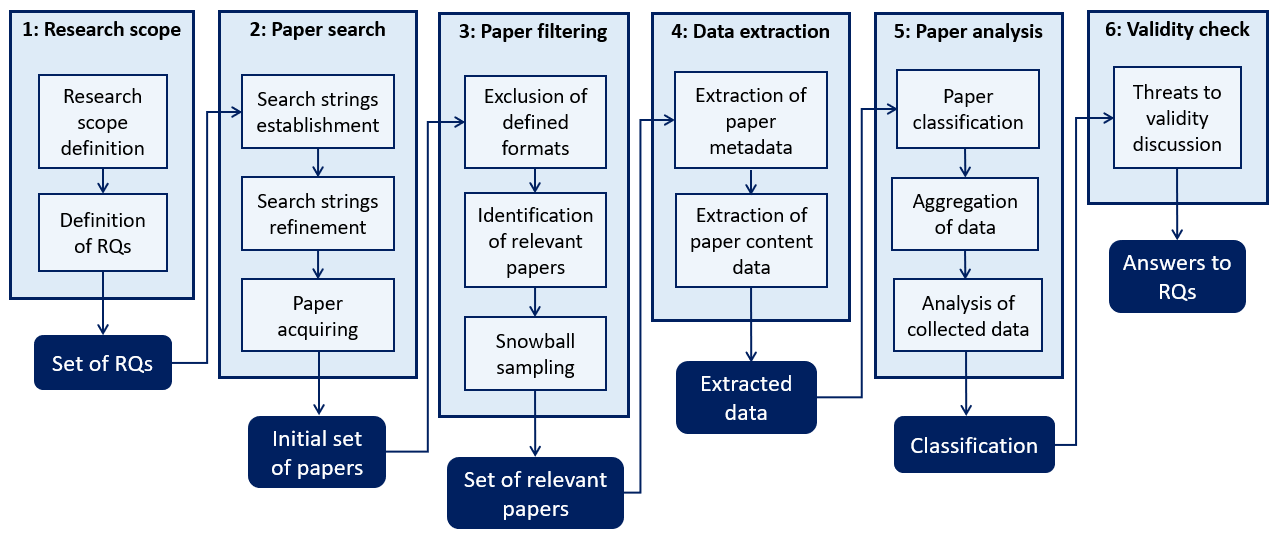}
  \caption{Stages of the systematic mapping study methodology followed in this study}
  \label{fig:methodology_process}
\end{figure*}

The main stages of the methodology are depicted in Fig. \ref{fig:methodology_process} and described in this section.

In this study, we define seven RQs for analyzing the field of integration testing methods for IoT systems from various viewpoints:
\begin{itemize}
\item \textbf{RQ 1:} What is the research trend in this field in terms of the number of studies published in recent years?
\item \textbf{RQ 2:} Which researchers are currently conducting active research in this field?
\item \textbf{RQ 3:} Which publication media (journals and conferences) publish papers in the field of integration testing for IoT systems?
\item \textbf{RQ 4:} What are the topics and subproblems currently being dealt with in the field of IoT integration and interoperability testing?
\item \textbf{RQ 5:} Which testing techniques and approaches are used in this field?
\item \textbf{RQ 6:} What are the current challenges and limitations in the field of IoT integration testing?
\item \textbf{RQ 7:} What are the possible future research directions in this field?

\end{itemize}

We do not limit the study to a particular class or type of subproblems, or testing techniques. Hence, \textit{RQ 4} involves informal testing techniques as well as formal and MBT techniques.


To search for relevant papers in the field of integration testing for IoT systems, we decided to use the following four established publication databases: IEEE Xplore, ACM Digital Library, Springer Link, and Elsevier ScienceDirect.

To verify the completeness of the search strings, we randomly selected a set of 30 papers as control samples, which discussed interoperability and integration testing issues of IoT systems. These control papers had to be present in the set of papers found using the search strings.

After a couple of refinement cycles, a general search string was finally established as

\textit{('Integration Testing' AND IoT) OR ('Integration Testing' AND 'Internet of Things') OR ('Interoperability Testing' AND IoT) OR ('Interoperability Testing' AND 'Internet of Things')} 

where the expression in apostrophes denotes the exact string to be searched at the same time. The general search string has been adopted based on particular notations used by individual databases. The timespan was determined to be from 2009 to 2019.

Journal papers, book chapters, and conference papers were selected for download. In the initial stage, we also downloaded conference posters and popular magazine articles, which were subsequently filtered. The number of initially downloaded papers is presented in Table \ref{table:number_of_papers_filtering_snowball}, column \textit{Initial sample size}.

Once the papers were downloaded, they were filtered in several steps. First, we excluded conference posters, papers shorter than two pages, and articles from popular magazines. Subsequently, two members of our research lab independently analyzed the paper title, abstract, and full text to assess whether the downloaded papers were relevant to the examined field. This process was conducted in parallel and the results were compared; in the case of mismatch of results, the relevance of the paper was assessed in a discussion until a consensus was reached. This was the case for 11\% of the analyzed studies on average. The number of filtered papers for individual databases is presented in Table \ref{table:number_of_papers_filtering_snowball}, column \textit{After filtering}.

In the next step, we followed the snowball sampling process; here, we analyzed other relevant papers and articles found in the references of the filtered papers, which were not already a part of the set of filtered papers. Studies and reports found during this sampling underwent the same filtering and assessment process as the downloaded set of papers; two lab members independently analyzed the title, abstract, and full text of the papers. 

The majority of the papers acquired by the snowballing process were obtained from the four major databases employed in this study (IEEE Xplore, ACM Digital Library, Springer Link, and Elsevier ScienceDirect) and two papers have been obtained from other databases. The described filtering process has been applied to the papers acquired by snowballing regardless of their source database.

Those papers that were found relevant were added to the analyzed sample. The number of papers found by individual databases after this step is presented in \ref{table:number_of_papers_filtering_snowball}, column \textit{After snowball}.


\begin{table}
\caption{Numbers of papers after filtering and snowball sampling.}\label{table:number_of_papers_filtering_snowball}
\begin{tabular}{|p{3cm}|c|c|c|}
\hline
\bfseries Source & \bfseries Initial sample size & \bfseries After filtering & \bfseries After snowball\\
\hline
IEEE Xplore & 384 & 45 & 53\\
\hline
ACM Digital Library & 87 & 10 & 12\\
\hline
Springer Link & 199 & 32 & 32\\
\hline
ScienceDirect & 133 & 15 & 16\\
\hline
other databases & 0 & 0 & 2\\
\hline
\bfseries total & \bfseries 803 & \bfseries 102 & \bfseries 115\\
\hline
\end{tabular}
\end{table}





During the data extraction and analysis phase, extracted data were independently verified by a specialist, who analyzed the set of papers and matched them with extracted metadata. A "two pair of eyes" approach was adopted for paper classification. Two specialists classified the papers independently; in the case of a mismatch, particular cases were discussed, papers were analyzed, and the final decision was made based on the discussion results. During this analysis, 8\% of the papers underwent mentioned discussion because of mismatch in the classification. The final set after this phase contained 115 papers.

\begin{table*}[!t]
\renewcommand{\arraystretch}{1.3}
\caption{Categories used in the paper classification}
\label{table:paper_classification}
\centering
\begin{tabular}{|p{2cm}|p{2.5cm}|p{7.6cm}|}
\hline
\bfseries Main category & \bfseries Category short name & \bfseries Category description  
\\
\hline\hline
- & IoT quality discussion & Interoperability/integration testing included in a general IoT quality discussion 
\\
\hline
\multirow{6}{2cm}{Testing methodology} & Testing methodology, including & General testing methodology including interoperability/integration testing as its part 
\\
\cline{2-3}
& Focused testing methodology
 & Methodology specially focused on interoperability/integration testing 
 \\
\cline{2-3}

 & Formal techniques & Formal testing/verification techniques for interoperability/integration of IoT systems 
 \\
\cline{2-3}
 & Testing methodology, applicable & General testing methodology applicable to interoperability/integration testing 
 \\
 \cline{2-3}
 & Literature review & Literature review related to IoT testing methods, which also includes interoperability and integration aspects 
 \\

\hline

\multirow{6}{2cm}{Testing frameworks, tools and testbeds} & Testing frameworks, supporting & General testing framework directly supporting interoperability/integration testing 
\\
\cline{2-3}
 & Test automation framework & Specialized test automation framework directly supporting integration testing 
 \\
\cline{2-3}
 & Testbeds & Report on IoT testbed directly supporting interoperability/integration testing 
 \\
\cline{2-3}
 & Testing framework, applicable & General testing framework applicable to interoperability/integration testing 
 \\
  \cline{2-3}

 & Frameworks and tools overview & Overview of testing frameworks and tools applicable to integration testing 
 \\
\hline

\multirow{3}{2cm}{Simulation frameworks} & Simulation frameworks, applicable & General IoT simulation frameworks applicable to interoperability/integration testing 
\\
  \cline{2-3}
    & Simulation frameworks, supporting & General IoT simulation framework supporting interoperability/integration testing 
    \\
\hline

- & Development frameworks & IoT systems development framework / approach / standard including interoperability/integration testing 
\\
\hline
\end{tabular}
\end{table*}

The narrowed selection of the papers was analyzed by publication year to answer RQ1 and by author names and affiliations to answer RQ2. Publication media were categorized by type (journal article, book chapter, conference paper, and workshop paper) and name to answer RQ3. Then, to answer RQ4, we classified the papers by categories presented in Table \ref{table:paper_classification}. Categories were organized in two levels: main category and subcategories. Subsequently, full text and detailed analysis of the paper content were used to answer RQs 5 to 7. 

The final set of 115 papers with their metadata including abstract, category, source URL, source library, and BibTex string are available for download at
\textit{http://still.felk.cvut.cz/iot\_integration\_testing/}. In the folder, the list is available in CSV, OpenOffice spreadsheet, and MS Excel format.


\section{Results}
\label{sec:results}

This section presents the results of the conducted analyses and answers to the individual RQs. Answers to each RQ are provided in a separate subsection.

\subsection{RQ1: Publication Trend in Time}

\begin{figure}[htbp]
\centering
  \includegraphics[width=8.5cm] {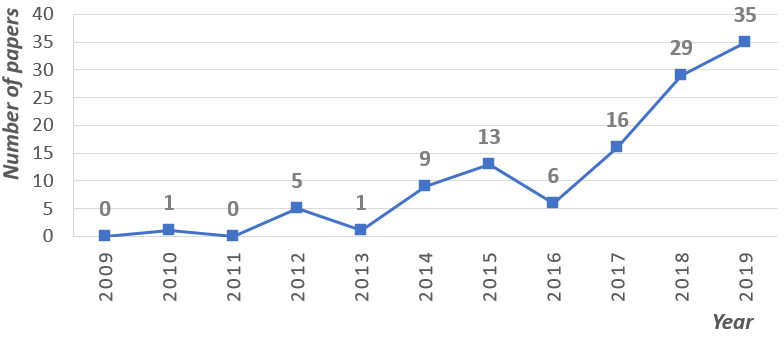}
  \caption{Number of publications by years}
  \label{fig:publications_by_years}
\end{figure}

In the recent decade, the number of publications discussing interoperability and integration testing issues of IoT systems has constantly grown, as shown by the data presented in Figure \ref{fig:publications_by_years}.

A more significant number of publications started to appear since 2014. The growth in publication numbers from 2016 to 2019 is almost constant. When extrapolating the trend, we can expect similar growth of publications discussing interoperability and integration testing of IoT systems in the following years.


\subsection{RQ2: Active Researchers}

In the final set of analyzed relevant studies, eight authors emerged to be actively publishing in the field of interoperability and integration testing of IoT systems. They were Brian Pickering (University of Southampton, UK),
Bruno Lima (University of Porto, Portugal),
Hamza Baqa (Institut Polytechnique Paris, France),
Koray Incki (Ozyegin University, Turkey),
Mengxuan Zhao (France Telecom),
Michael Felderer (University of Innsbruck, Austria),
Paul Grace (Aston University, UK), and
Thomas Watteyne (Inria, France); they all published three studies.

No author from the analyzed set published more than three studies from 2009 to 2019, 29 authors published two studies, and 431 authors published one study. A total of 468 unique authors were found in the analyzed studies.

This analysis also points out the relative heterogeneity of the research community and absence of research mainstream in this field. However, this is a contemporary situation and might change in the near future.


\subsection{RQ3: Publication Media in IoT Integration Testing}

\begin{figure*}[htbp]
\centering
  \includegraphics[width=12cm] {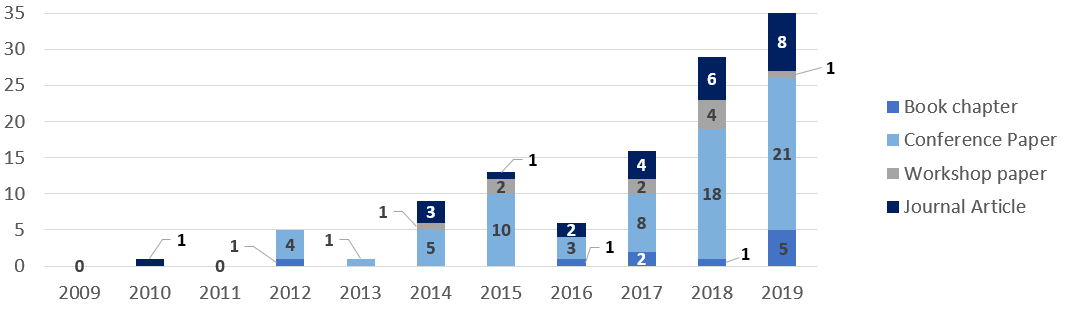}
  \caption{Venue types by individual years}
  \label{fig:venue_types_by_years}
\end{figure*}

During the analysis of the papers, we analyzed four main publication media types: journal article, conference paper, workshop paper, and book chapter. Papers of conference proceedings published in a book series (e.g., LNCS by Springer) were considered as conference papers. Among the analyzed set of papers, several have been published in conferences aggregating parallel workshops; such papers were also considered as conference papers. Most papers were published in conference proceedings (61\%), followed by journal articles (22\%), workshop papers (9\%), and book chapters (9\%). Figure \ref{fig:venue_types_by_years} presents more details on the publication media type by individual years of the analyzed period.


Among the media types of the published studies on interoperability and integration testing, we will start by analyzing conference papers. The analyzed studies were published in a wide variety of conferences spanning from established conferences in system testing (e.g., IEEE Conference on Software Testing, Validation and Verification (ICST), IFIP International Conference on Testing Software and Systems, and IEEE International Conference on Software Quality, Reliability and Security (QRS)) to various forums related to IoT technology (e.g., IEEE World Forum on Internet of Things (WF-IoT), IEEE International Conference on Future Internet of Things and Cloud Workshops (FiCloudW), and European Conference on Wireless Sensor Networks).

However, the spectrum of the conferences publishing papers focusing on IoT integration and interoperability testing is rather heterogenic, and apart from a few exceptions, we have not found a leading conference publishing more than three papers in the analyzed sample. The IEEE World Forum on Internet of Things (WF-IoT) published three papers, and the Global Internet of Things Summit (GIoTS) published two papers. The remainder of the analyzed papers were published in various unique conferences.

Regarding the journals publishing IoT interoperability and integration testing studies, the situation was found to be similar. Articles were published in a relatively wide spectrum of journals dedicated to computer systems, networks, software testing, and related areas. Three articles were published in IEEE Access and two articles in the International Journal on Software Tools for Technology Transfer. The remaining studies were published in various unique journals. The details can be found in the complete list of analyzed papers available at 

\textit{http://still.felk.cvut.cz/iot-integration-testing/}.

To summarize, publication media for integration and interoperability testing studies are relatively heterogenic. Even though integration and interoperability testing are understood as established discipline in the industrial praxis, in the research world, no major journal or conferences outlies as a venue especially publishing in this specific field. This can be explained by the relative novelty of the field. However, considering that the present industry calls for more effective and systematic methods for interoperability and integration testing, the research community will very likely react to these demands, and the situation will possibly change in the coming years.

\subsection{RQ4: Topics and Subproblems being Addressed}

\begin{figure*}[htbp]
\centering
  \includegraphics[width=12cm] {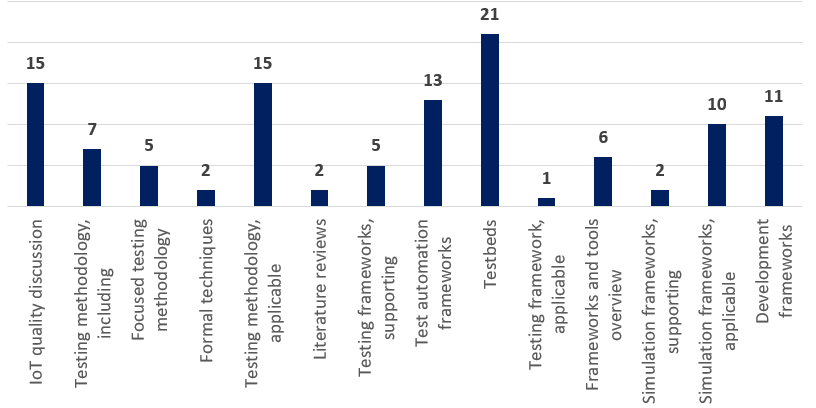}
  \caption{Classification of analyzed studies}
  \label{fig:classification}
\end{figure*}

Figure \ref{fig:classification} presents the classification of analyzed relevant studies using the categories shown in Table \ref{table:paper_classification}. The complete list of individual papers assigned to each category are given in the link above. In the analyzed sample, two major groups were found to be testing methodologies supporting or related to interoperability and integration testing (main category \textit{Testing methodology} with 31 papers in total); and testing frameworks and testing tools, including test automation tools and testbeds constructed for or supporting interoperability and integration testing of IoT systems (main category \textit{Testing frameworks, tools, and testbeds} with 46 papers in total). The analyzed set of papers also includes various IoT simulation frameworks applied to IoT interoperability and integration testing (main category \textit{Simulation frameworks} with 12 papers in total).

In the detailed categories, the largest number of analyzed studies include discussions on various IoT testbeds supporting integration testing (21 papers), followed by general IoT quality discussions (15 papers), and IoT testing methodologies applicable to integration and interoperability testing (15 papers). Interoperability and integration of IoT systems are discussed in 13 papers dedicated to IoT test automation frameworks. This topic is also the subject of 11 studies presenting various development frameworks for IoT solutions.

On the contrary, the presence of formal methods in the analyzed papers is low; only two papers focus on this topic. Similarly, only five studies present a directly focused integration testing methodology. We analyze the used techniques and approaches in section \ref{sec:testing_techniques}.

Two of the analyzed papers were also literature reviews relevant to the scope of this paper: a literature review dedicated to testing methods for embedded systems \cite{Garousi_2018} and a study summarizing general testing methods for the IoT filed \cite{10.1145/3356317.3356326}.

\subsection{RQ5: Used Testing Techniques and Approaches}
\label{sec:testing_techniques}

In the studies relevant to interoperability and integration testing of IoT systems, a variety of testing techniques and approaches have been researched and applied, from formal verifications \cite{Camilli_2018} to informal crowdsourcing techniques that can be compared to exploratory testing \cite{7104907}.

In this study, by a testing approach, we mean (1) general approach to test design and test execution, e.g., formal verifications, Model-based Testing or informal testing, and, (2) generic testing approaches based on various test levels as unit testing, integration testing or acceptance testing, for instance.
By testing techniques, we mean techniques to create test cases; for instance, combinatorial or constrained interaction testing, path-based testing, and data-flow testing.

Regarding established testing techniques, path-based testing using finite state machines (or analogous structures) as a system under test (SUT) model is discussed in five studies \cite{6294224,Amalfitano:2017:TTA:3137003.3137007,10.1007/978-3-030-31280-0_1,Grace_2015,7579811}. In addition, an SUT model based on a timed state machine has also been employed \cite{Larsen_2017}.

Datta et al. presented a prospective approach to semantic interoperability testing of IoT devices or services \cite{Datta_2018}. In their concept, they distinguish between syntactic and semantic interoperability to be verified during the tests. Semantic testing is also employed in a test automation framework proposed by Kim et al. \cite{Kim_2018}.

Regarding the established test case notations, TTCN-3 standard by ETSI has been employed in six proposals \cite{Schieferdecker_2017,Park_2016,Rings_2013,6294224,Makedonski_2018,Sotiriadis_2017}.

Nevertheless, established testing techniques related to IoT integration testing in the studied papers are few. In contrast, general testing approaches are discussed more intensely.

The MBT approach, in general, is explicitly discussed in several studies \cite{ahmad2018model,Ahmad_2016,Incki_2018_2,267044762,10.1007/978-3-030-31280-0_1,Lima_2019,Grace_2015,7579811}, which mostly describe a general concept; particular testing technique, namely path-based testing, is discussed in the studies by Aicherning et al. \cite{10.1007/978-3-030-31280-0_1}, Estivill-Castro et al. \cite{7579811},
 and Grace et al. \cite{Grace_2015}. 

Suggestions of formal verifications \cite{Camilli_2018} and runtime verifications \cite{Incki_2018} do appear; however, for integration and interoperability testing of IoT systems, these have to be further elaborated.

Mutation testing has been used by Lu et al. for verification of RFID devices \cite{Lu_2010}; this technique can be expected to be used in future works to verify the effectiveness of developed testing techniques. 

Other testing approaches include use case testing \cite{10.1145/3350768.3350796,10.1145/3364641.3364644}, and, practically, exploratory testing and error guessing \cite{7104907}.

Several studies suggest test strategies and approaches for IoT systems that consist of general test levels (e.g., unit testing, integration testing, and acceptance testing) and approaching these test levels informally (e.g., testing of individual sensors, testing or integration, and security testing) \cite{8365717,Sand_2016,Pontes:2018:IPI:3236454.3236511,Eckhart_2019,Tan_2019,Pontes:2018:TPI:3278186.3278196,9043692,9014711,8859486,Felderer2019,WALKER2019155}. These studies can be used as a basis for setting up a test strategy for an IoT system.

\subsection{RQ6: Challenges and Limitations}

After the analysis of the current studies in terms of interoperability and integration testing and comparing the state-of-the-art methods with the current industry demand, several conclusions can be drawn and several prospective research directions can be suggested.

The research community in the interoperability and integration testing of IoT systems seems rather heterogenic; from the analyzed studies, there is no clear leading publishing medium or author in this field focusing on this topic. This heterogeneity can be explained by a combination of several factors:
\begin{enumerate}
\item The field of IoT testing and quality assurance is relatively novel; despite the active production of innovative IoT solutions in the last five years, research and development of IoT-specific testing methods is currently a developing field.
\item General methods from the field of integration testing might be considered as satisfactory for testing IoT systems; thus, demand for IoT-specific interoperability and integration testing methods is not specially recognized in the research community.
\item In the research community, several research streams and subcommunities have been established, covering path-based testing, combinatorial interaction testing, constrained interaction testing, data-flow testing, and other individual basic testing techniques that can be combined to establish comprehensive integration testing methods. Hence, interoperability and integration testing itself is not considered as a subject of primary research. Instead, the focus is on primary testing approaches that can be employed for the interoperability and integration testing process.
\end{enumerate}

In particular, the second and third points deserve further analysis and discussion. Regarding the second point (hypothetical low necessity to develop IoT-specific interoperability and integration testing methods, because there are general testing methods for these cases already available), it is worthwhile to analyze the situation in the current IoT systems briefly. Compared to standard software systems or relatively isolated proprietary cyber-physical systems not connected to the Internet, the situation in interconnected IoT systems might be different for a number of cases. In these systems, a more extensive set of various protocols on different networks and application levels can be integrated together, and seamless integration has to be maintained. These protocols might span from standardized protocols like WiFi, Bluetooth, IEEE 802.15.4, Z-wave, or ZigBee for low levels of the system; REST, MQTT, CoAP, or LWM2M protocols for higher levels of the system \cite{al2015internet}; to various proprietary protocols used by individual vendors. These proprietary protocols might also contain more defects than established standards, and this fact makes smooth interoperability and integration of an IoT system more challenging. 

This situation leads to the increased necessity to employ techniques testing correct functionality of integration interfaces and interoperability with different configurations. It also leads us to suggest that the current testing methods shall be revised in the IoT context to increase their potential to detect relevant defects.

The same applies to individual devices, where the level of standardization might be relatively low. Several attempts to standardize IoT devices and allow their interoperability have been made (e.g., ETSI GS CIM 009 or ISO/IEC 21823); however, no major standard is currently established. This is the reason for significant integration and interoperability challenges. 

Therefore, the capability of previous interoperability and integration testing techniques should be revised, at the minimum; opportunities to create more effective approaches based on IoT system specifics have to be examined. These opportunities cover Combinatorial and Constrained Interaction Testing \cite{ahmed2017constrained} as well as path-based testing and data-flow testing \cite{su2017survey} techniques for integration testing (typically end-to-end integration tests).

Regarding the third point, the argument that interoperability and integration testing itself might not be understood as a subject of primary research, rather as an application of primary testing approaches, there are two counter-arguments worth mentioning. First, in general system testing research, integration testing is understood as a standalone research topic, as is documented in previous mapping studies \cite{shashank2010systematic,haser2014software}; in particular, the study by Haser et al. documents the broad extent of studies dedicated to integration testing of software and cyber-physical systems \cite{haser2014software}.
Moreover, another finding by a recent study conducted by Ahmed et al. \cite{ahmed2019aspects} should be considered. Even in the discussed primary testing approaches such as path-based testing, combinatorial interaction testing, constrained interaction testing, or data-flow testing, no specific variants of these techniques are published for IoT systems to a large extent.

Hence, to summarize, relative heterogeneity of the IoT interoperability and integration testing approaches might be explained as a result of the relative novelty of the field. Further development of IoT-specific testing techniques to cover these areas is a prospective future research direction. We analyze potential research directions further in Section \ref{sec:future_research_directions}.




\subsection{RQ7: Future Research Directions}
\label{sec:future_research_directions}

Regarding interoperability and integration testing methods for IoT systems, several prospective future research directions can be discussed considering the industrial needs and specifics of IoT systems. 

First, specific techniques for integration and interoperability testing of IoT systems have not yet been studied extensively in the literature. The techniques might have been published under different names; for experts in the field, it might be an easy task to get an overall picture. However, for testing practitioners and researchers from other fields, getting such a picture might be more difficult.

The first future research area is handling possible combinatorial explosion problems in integration testing when considering possible configurations to test large-scale IoT systems. When various devices are integrated together in IoT systems, where these devices may vary in versions, many different system configurations can be established; flawless interoperability of devices in these variants need to be tested. The current combinatorial \cite{nie2011survey} and constrained \cite{ahmed2017constrained} testing disciplines handle the problem on a general level. However, IoT-specific support regarding the modelling of the problem and application of general combinatorial techniques to IoT and integration testing using specific metadata from an IoT system might represent another perspective direction for future research.

Another relevant field is testing the seamless integration of various devices in an IoT system operating with limited network connectivity. Transmission of data from sensors and between actuators operating in areas with weak network signal coverage might be disrupted during the system run. Hence, in such situations, the overall functionality of an IoT system should be checked for functional correctness and transaction processing of the data, if required. To the best of our knowledge, in the testing of such reliability, current publications focus on lower levels of the system (typically network layer), and systematic methods for such tests on higher levels of an IoT system have yet to be provided.

In addition, to ensure more effective tests and also give the testing practitioners better guidance on how to construct test cases, cross-over techniques between path-based testing \cite{anand2013orchestrated} and combinatorial interaction testing \cite{nie2011survey} for testing of close APIs in IoT systems might be researched. Using specific information and metadata from the tested system usually helps focus on the test cases more effectively, and this direction can also be explored in the case of IoT systems.

\section{Conclusion}

In this study, we focused on the field of integration and interoperability testing of IoT systems. The motivation was twofold: the importance of this field in the current industry and the fact that this specific area has not yet been covered by a focused, systematic literature mapping study.

In the study, we analyzed 803 papers from four major primary databases, namely, IEEE Xplore, ACM Digital Library, Springer Link, and Elsevier ScienceDirect and followed the current established recommendations for conducting mapping studies by Kitchenham and Charters \cite{kitchenham2007guidelines}. After a detailed assessment of the papers and quality check, 115 papers were found to be relevant to the field. 

Our results suggest that currently there are general testing methods, which can be applied to the field of integration and interoperability testing of IoT systems; therefore, there is a research opportunity to evolve more specific testing methods directly focused on IoT systems, which might work more effectively in the IoT context.

On the other hand, a number of testing and test automation frameworks that support interoperability and integration testing are being created already, and we can also find examples of individual testbeds supporting this field.

There may be several concerns related to the validity of this study. The main concern may be the exclusion of some relevant papers from the list. This possible problem was effectively mitigated by multiple-stage paper filtering and snowballing process, as described in Section \ref{sec:methodology}, which also includes a thorough validity check phase.

Another possible concern may be the inclusion of irrelevant papers in the scope, which was also mitigated by the methodology (see Section \ref{sec:methodology}) following well-known methods for the selection criteria as well as the "two pairs of eyes" quality check.

A limitation of this mapping study is that it analyzes papers published only in the four primary major databases (IEEE Xplore, ACM Digital Library, Springer Link and Elsevier ScienceDirect) and does not involve other possible sources such as Google Scholar, Scopus, researchgate.net or arxxiv.org, which might contain other relevant studies.

Despite these possible limitations, several prospective research directions were suggested in this study.

\textit{This research is conducted as a part of the project TACR TH02010296 Quality Assurance System for the Internet of Things Technology. The authors acknowledge the support of the OP VVV funded project CZ.02.1.01/0.0/0.0/16\_019 /0000765 “Research Center for Informatics”.}

%
%
%
\bibliographystyle{splncs04}
\bibliography{references}

\end{document}